\documentclass[journal]{IEEEtran}

\ifCLASSINFOpdf
\else
   \usepackage[dvips]{graphicx}
\fi

\usepackage{amsmath}
\usepackage{textcomp}

\usepackage{booktabs, multirow} % for borders and merged ranges
\usepackage{soul}% for underlines
\usepackage[table]{xcolor} % for cell colors
\usepackage{changepage,threeparttable} % for wide tables
\usepackage{hyperref}
\usepackage{tablefootnote}

\usepackage{graphicx}

\begin{document}

\title{A Data-Driven Analysis of \\ Robust Automatic Piano Transcription}

\author{Drew Edwards, Simon Dixon, Emmanouil Benetos, Akira Maezawa, Yuta Kusaka
%\thanks{This work was supported in part by UK Research and Innovation under Grant EP/S022694/1. EB is supported by RAEng/Leverhulme Trust Research Fellowship LTRF2223-19-106.}
%\thanks{Drew Edwards, Simon Dixon, and Emmanouil Benetos are with Queen Mary University of London, London, E1 4NS UK (a.c.edwards@qmul.ac.uk).}
%\thanks{Akira Maezawa and Yuta Kusaka are with Yamaha Corporation, Hamamatsu, Shizuoka 430-8650, Japan.}
\thanks{DE, SD, and EB are with Queen Mary University of London (QMUL), London, E1 4NS, UK (a.c.edwards@qmul.ac.uk); AM and YK are with Yamaha Corporation, Hamamatsu, Shizuoka 430-8650, Japan. DE is a PhD student at the UKRI CDT in AI and Music, supported by UKRI (EP/S022694/1), QMUL, and Yamaha. EB is supported by RAEng/Leverhulme Trust Research Fellowship (LTRF2223-19-106). This project used the Tier 2 HPC facility JADE2, funded by EPSRC (EP/T022205/1). For the purpose of open access, the authors have applied a Creative Commons Attribution (CC BY) license to any Accepted Manuscript version arising.}
}

%\author{First A. Author, \IEEEmembership{Fellow, IEEE}, Second B. Author, and Third C. Author, Jr., \IEEEmembership{Member, IEEE}
%\thanks{This paragraph of the first footnote will contain the date on which you submitted your paper for review. It will also contain support information, including sponsor and financial support acknowledgment. For example, ``This work was supported in part by the U.S. Department of Commerce under Grant BS123456.'' }
%\thanks{The next few paragraphs should contain the authors' current affiliations, including current address and e-mail. For example, F. A. Author is with the National Institute of Standards and Technology, Boulder, CO 80305 USA (e-mail: author@boulder.nist.gov).}
%\thanks{S. B. Author, Jr., was with Rice University, Houston, TX 77005 USA. He is now with the Department of Physics, Colorado State University, Fort Collins, CO 80523 USA (e-mail: author@lamar.colostate.edu).}}

\markboth{Journal of \LaTeX\ Class Files, Vol. 14, No. 8, August 2015}
{Shell \MakeLowercase{\textit{et al.}}: Bare Demo of IEEEtran.cls for IEEE Journals}
\maketitle

\begin{abstract}
Algorithms for automatic piano transcription have improved dramatically in recent years due to new datasets and modeling techniques. Recent developments have focused primarily on adapting new neural network architectures, such as the Transformer and Perceiver, in order to yield more accurate systems. In this work, we study transcription systems from the perspective of their training data. By measuring their performance on out-of-distribution annotated piano data, we show how these models can severely overfit to acoustic properties of the training data. We create a new set of audio for the MAESTRO dataset, captured automatically in a professional studio recording environment via Yamaha Disklavier playback. Using various data augmentation techniques when training with the original and re-performed versions of the MAESTRO dataset, we achieve state-of-the-art note-onset accuracy of 88.4 F1-score on the MAPS dataset, without seeing any of its training data. We subsequently analyze these data augmentation techniques in a series of ablation studies to better understand their influence on the resulting models.
\end{abstract}


\IEEEpeerreviewmaketitle

\section{Introduction}
\label{sec:intro}
\IEEEPARstart{A}{utomatic} music transcription (AMT) is the signal processing task of converting an audio recording of a musical performance into symbolic form. The target output could be a low-level physical description of each note played by any instrument in the recording, such as MIDI, or sheet music with representations of musical features such as meter, time signature, rhythm, instrumentation, and dynamics. 

In this work, we focus on AMT applied to piano, or automatic piano transcription (APT). APT has improved dramatically over the past decade. The two largest relative improvements in state-of-the-art accuracy were from a new architecture to jointly predict onsets and frames \cite{onsetsandframes} and from the release of the MAESTRO dataset \cite{maestro}. The Onsets and Frames architecture yielded a more than fifty percent increase in the note-onset F1-score, when trained and evaluated on the MAPS dataset \cite{maps}. The MAESTRO dataset yielded an additional five percent relative improvement on MAPS test evaluation in a zero-shot setting, along with becoming the de-facto standard in training APT systems. Since then, the research has largely focused on new model architectures and adapting these systems for multi-instrument transcription \cite{kong_high_resolution, gardner2022mt, Lu2023, toyama2023}.

However, most of this research ignores out-of-distribution\footnote{In the context of this research, ``out-of-distribution'' refers to solo piano recordings drawn from a different distribution than the training data, e.g. with different pianos, recording conditions, pieces and musical styles.} performance of their trained models. The emphasis has been on state-of-the-art results on held out test sets, despite the well-documented phenomenon of overfitting to acoustic conditions shared between test and training sets \cite{albumeffect}. This lack of focus on generalization can lead to models which exhibit shortcut learning \cite{Geirhos2020}, and thus the reported evaluation metrics can overstate the true performance of these systems. Furthermore, publications may forgo techniques to make their models more robust and generalizable, as this has been demonstrated to slightly hurt their test set results. This is in contrast to computer vision literature where data augmentation has become standard practice in training deep learning models \cite{Shorten2019}.

In this research, we analyze the robustness of APT with a data-driven methodology. We demonstrate how existing systems can overfit to the acoustic conditions of their training data, which can hamper generalization. We then perform a series of experiments to analyze the impact of data augmentation on out-of-distribution transcription performance. Our core contributions are the following:
\begin{itemize}
\item a detailed analysis of data augmentation techniques applied to the task of APT;
\item an automatically re-performed version of all $\approx 200$ hours of the MAESTRO dataset captured in a studio recording environment via Yamaha Disklavier technology;
\item showing how a combination of data augmentation and increased timbral diversity results in state-of-the-art (SOTA) performance on the MAPS dataset, and making the model weights publicly available.
\end{itemize}

\begin{figure*}[t]
\includegraphics[width=0.95\textwidth]{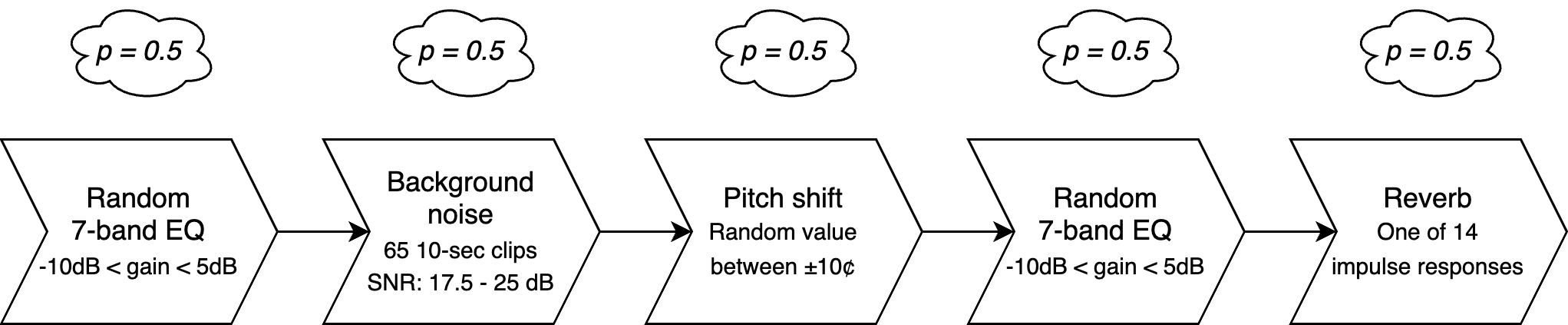}
\centering
\caption{Illustration of the data augmentation pipeline employed. The process flows from left to right, starting with a Random 7-Band EQ, where each band's gain is randomly adjusted within a range of -10dB to 5dB. Next, random background noise is added, selected from a set of 65 different 10-second clips, with the signal-to-noise ratio meticulously maintained between 17.5 and 25dB. Then we apply a random pitch shift, varying between -10 and 10 cents, to account for slight tuning discrepancies in production data. Another Random 7-Band EQ is then applied. The final stage of the pipeline introduces reverb, utilizing one of 14 unique impulse responses, to simulate different spatial acoustic characteristics. Each component of this pipeline is applied independently with a probability of 0.5, ensuring a rich and diverse set of augmentations. \label{fig:data_aug_pipeline}}
\end{figure*}

\section{Related Work}
\label{sec:related}
Deep learning models have become the current SOTA in AMT. The first end-to-end neural network approach for piano transcription was by Sigtia et al.\ \cite{e2epiano}. This was soon improved by Hawthorne et al.\ \cite{onsetsandframes} by formulating a dual-objective of predicting both note onsets and frame activity. Recognizing that their approach was limited by data, the same authors introduced the MAESTRO dataset \cite{maestro}: 200 hours of virtuoso solo piano performances with finely-aligned MIDI labels. Before MAESTRO, the largest available dataset for piano transcription was MAPS \cite{maps}, which has 18 hours of audio. Soon after the MAESTRO publication, Kong et al.\ \cite{kong_high_resolution} introduced a novel technique to regress on the onset times, allowing higher resolution than the spectrogram frame width. Subsequent attempts have introduced the Transformer \cite{seq2seq,gardner2022mt} and Perceiver \cite{Lu2023} architectures for piano and multi-instrument transcription. The current SOTA is from the recent work of Toyama et al., which relies upon the self-attention mechanism of Transformers to capture long-term dependencies in the frequency and time axes, yielding a note-level transcription F1 score of 97.4 evaluated on the MAESTRO test set \cite{toyama2023}. 

The main technique we explore for regularizing deep transcription models is data augmentation. There has been much work on using data augmentation for audio based learning tasks \cite{salamon2017deep, mauch2013audio, maestro, McFee2015ASF}. Thickstun et al.\ \cite{thickstun2018invariances} explore label-preserving pitch-shift transformations. Simon et al.\ \cite{scalingpolyphonictranscription} use mixtures of monophonic transcription examples to pre-train their multi-instrument polyphonic transcription transformer. Lu et al.\ \cite{Lu2023} use cross-dataset mixtures for training, along with label-preserving pitch shifts. The \texttt{audiomentations} library\footnote{\href{https://github.com/iver56/audiomentations}{https://github.com/iver56/audiomentations}} provides a suite of audio-based data augmentations to apply on the fly during machine learning training, which we make use of in this work.

\subsection{Robustness of Piano Transcription Systems}

Despite the great progress on this task, nearly all of the results mentioned above fail to report out-of-distribution evaluation metrics. The one exception is the MAESTRO paper \cite{maestro}, which emphasises the importance of using data augmentation when evaluating on the MAPS test set. A very promising direction in robust transcription research comes from Maman and Bermano \cite{maman}: the authors describe a technique to automatically align scores with live performances. Using only synthetic data and automatically aligned scores, they achieve the current SOTA on the MAPS test set and report very encouraging cross-dataset performance. 

Many factors seem to contribute to a more robust piano transcription system. First, certain data augmentation techniques can improve generalization. For example, models can overfit to the tuning of pianos in their training data, and small amounts of random pitch shifting can mitigate this effect. Data augmentation is the core technique explored in this publication and is explored in detail in Section \ref{sec:aug}. Second, timbral and acoustic diversity seems to improve generalization. This is likely the explanation for the very robust performance of Maman and Bermano, since they train on MusicNet \cite{musicnet}, which includes audio from a wide variety of sources and includes multi-instrumental performances. Robustness to noisy training data is another factor: Kong et al.\ \cite{kong_high_resolution} demonstrate that their triangulation technique permits learning when note-onset labels are off by 50ms, while the architecture from Onsets and Frames is susceptible to much lower accuracy in this regime. 

\section{Studio MAESTRO Dataset}
\label{sec:studio}
The audio in the MAESTRO dataset contains recordings from ten years of piano competitions. All pieces are performed by a human on a Yamaha Disklavier piano in a concert hall with an audience present. While the instrument used in the various years may be the same, the tuning will certainly vary year to year. Furthermore the exact placement of microphones will not be identical. On one hand, this is desirable: the data is very realistic and contains useful noise and variations for better generalization. On the other hand, we may seek to remove these differences to better understand the influence of data augmentation and the extent of overfitting that state-of-the-art transcription models exhibit. 

To that end, we create a data collection apparatus to automate the recording of acoustic piano performances. The equipment used consists of a Yamaha C6X grand piano manufactured in 2008 with a Disklavier ENSPIRE PRO player piano controller unit, two Neumann U87Ai large dual-diaphragm microphones, two Schoeps MK 6 capsule microphones, an iPhone 12 Pro, a Fireface UFX audio interface, and a MacBook Pro. A Python tool was developed to playback MIDI through the Yamaha Disklavier while simultaneously maintaining an audio input stream. The timing of the MIDI playback is synchronized to the clock of the audio interface to prevent drift. The software used for data collection is made available on GitHub \footnote{\href{https://github.com/almostimplemented/piano-capture}{https://github.com/almostimplemented/piano-capture}}. The apparatus was used to re-record all 200 hours of the MAESTRO dataset, which is released as part of this publication and is available on Zenodo \footnote{\href{https://zenodo.org/records/10082144}{https://zenodo.org/records/10082144}}.

\section{Training Experiments}
\label{sec:aug}

All experiments in this paper use Kong et al.'s \cite{kong_high_resolution} model architecture and implementation. They use a deep neural network architecture that combines convolutional layers for feature extraction from input spectrograms with gated recurrent units to capture temporal dependencies. Precise onset and offset time predictions are achieved via regression heads. However, this choice is somewhat arbitrary: other powerful transcription models could have been used. Instead of focusing on modifications to the model architecture, we analyze various training conditions characterized by the training data.

\subsection{Data Augmentation Experiment}

Training with the data collected from the studio recording setup exhibited severe overfitting. Our initial training experiment used only the studio audio with no data augmentation. We evaluate on the test split of MAESTRO, first using the Studio MAESTRO audio and then using the original audio. On the Studio MAESTRO data, the note-onset metrics \cite{Bay2009EvaluationOM} are ($\mathit{Precision} = 99.75$, $\mathit{Recall} = 95.09$, $\mathit{F1} = 97.32$). On the original MAESTRO data, the performance falls to ($\mathit{Precision} = 77.08$, $\mathit{Recall} = 85.69$, $\mathit{F1} = 80.77$). We believe the lack of acoustic and timbral variety limits the ability of the network to generalize to new pianos. To address these shortcomings, we deploy a novel data augmentation pipeline and supplement the training data with the original MAESTRO audio plus six additional audio renderings (varying in timbre) of the MIDI by Modartt Pianoteq version 7 \footnote{\href{https://www.modartt.com/pianoteq_overview}{https://www.modartt.com/pianoteq\_overview}}.

Figure \ref{fig:data_aug_pipeline} shows the general scheme of our data augmentation pipeline, which is implemented with the \texttt{audiomentations} package and is inspired by the pipeline of Hawthorne et al.\ \cite{maestro} with a few modifications. We apply two random seven band parametric equalizers, additive background noise with a variable signal-to-noise (SNR) ratio, random pitch shifting (between $\pm 0.1$ semitones), and reverb. The background noise is inspired by the pub noise of the Audio Degradation Toolbox (ADT) \cite{mauch2013audio}: the audio files consist of recordings of noisy public environments, typically at cafes and bars. We use \texttt{restaurant08.wav} from ADT plus four recordings available on \href{https://freesound.org}{freesound.org}, cut into 10-second segments to apply at training time (each training example is also 10 seconds). The reverb is similarly generated from 14 impulse responses from \href{https://echothief.com}{echothief.com}. For each training example, there is a 50\% chance that each stage of the pipeline will be applied.

The neural network model is taken from Kong et al.\ \cite{kong_high_resolution}. During training we use a batch size of 32, a learning rate of $5 \times 10^{-4}$, and run for 200000 steps. Since we have 8 different sources of audio (original MAESTRO, Studio MAESTRO, and the six synthesized versions), we apply a sampling scheme of $1/4, 1/4, 1/12, ..., 1/12$ respectively. We forgo training the pedal predictor and focus on note-onset prediction, which is independent of the model's pedal predictions.

\subsection{Results}\label{sec:results}

We present results of evaluation on the MAPS test set configuration  \cite{e2epiano}. 
Note-onset precision, recall, and F1 are reported in Table \ref{tab:maps_eval}. 
In all cases, the models were not trained on the MAPS dataset. 
%We give two results for Kong et al.\. \cite{kong_high_resolution}: one with their published thresholds and one with thresholds that we selected for our version of their model trained with data augmentation. 
Maman and Bermano \cite{maman} was the existing SOTA on the MAPS test set, which can be attributed to the diversity of their training data. Toyama et al.\ \cite{toyama2023} hold the SOTA on the MAESTRO test set ($Precision = 99.64$, $Recall = 95.44$, $F1 = 97.44$), but their model fails to reach comparable performance in the out-of-distribution setting on MAPS. Our model achieves a new SOTA for MAPS, without ever training on the MAPS train split. The model architecture is identical to Kong et al.\ \cite{kong_high_resolution}, so the 6\% improvement owes exclusively to our data-driven methodology. It is worth noting that all models (except Maman and Bermano) suffer a considerable drop in performance when evaluating on the MAPS test set compared to that of MAESTRO. This motivates further investigation into regularization techniques of these models. In the next section, we look at the effect of our chosen data augmentations in finer detail. 

\begin{table}[!t]\centering
\caption{Out-of-distribution performance of state-of-the-art piano transcription models evaluated on the MAPS test set. }\label{tab:maps_eval}
\begin{tabular}{lrrrr}\toprule
&\multicolumn{3}{c}{Note-level metrics} \\\cmidrule{2-4}
Model&Precision &Recall &F1 \\\midrule

Hawthorne et al.\ \cite{maestro} &87.5 &85.6 &86.4  \\
Kong et al.\ \cite{kong_high_resolution}  &78.3 & 87.2 & 82.4 \\
%Kong et al.\ \cite{kong_high_resolution} {\footnotesize(high threshold)}  & 85.9 & 84.7 & 85.2 \\
Maman and Bermano \cite{maman} &88.2 &86.5 & 87.3 \\
Toyama et al.\ \cite{toyama2023} &84.6 &85.7 &85.1 \\
Ours &\textbf{89.5} & \textbf{87.4} &\textbf{88.4}\\
\bottomrule
\end{tabular}
\end{table}

\section{Data Augmentation Analysis}
We present several experiments to explore the effect of data augmentation on training piano transcription systems. The first experiment (\ref{sec:data_degradation_study}) applies our data augmentations independently at test time to demonstrate the sensitivity of a model trained without augmentation. The second experiment (\ref{sec:single_augmentations}) applies each augmentation at training time and measures the impact on out-of-distribution evaluation. The final experiment (\ref{sec:ablation_study}) is an ablation study, again at training time, that removes one component of the augmentation pipeline at a time.

\begin{threeparttable}[h]%\centering
\caption{{\bf Data Degradation Results}, showing the effect of data augmentation on performance. Kong et al.'s model is trained only on MAESTRO's training data. Our model benefits from a diversified training set, including original, studio, and synthesized MAESTRO recordings, all augmented. Evaluation spans test splits from MAESTRO, Studio MAESTRO, and MAPS, focusing on the note-onset F1 score.}\label{tab:data_degradation}
\scriptsize
\begin{tabular}{lrrrrrrr}\toprule
&\multicolumn{2}{c}{MAESTRO} &\multicolumn{2}{c}{Studio} &\multicolumn{2}{c}{MAPS} \\\cmidrule{2-7}
&Kong \cite{kong_high_resolution} &Ours & Kong \cite{kong_high_resolution} &Ours & Kong \cite{kong_high_resolution} &Ours \\\midrule
No Augmentation &96.8 &96.6 &95.7 &97.7 &82.4 &88.4 \\
Background Noise &93.4 &95.4 &91.4 &97.1 &82.6 &88.1 \\
EQ &96.6 &96.5 &95.1 &97.7 &81.8 &88.4 \\
Pitch Shift &85.2 &94.1 &76.5 &96.0 &72.4 &87.3 \\
Reverb &89.1 &95.2 &87.0 &96.9 &72.3 &87.1 \\
\bottomrule
\end{tabular}
\end{threeparttable}%\end{adjustwidth}

\subsection{Data Degradation Study}
\label{sec:data_degradation_study}
As an initial investigation of the robustness of published models, we show in Table \ref{tab:data_degradation} how simple augmentations to test data can cause a dramatic drop in accuracy. We decompose our data augmentation pipeline and apply each augmentation in isolation during test set evaluation. We perform this on three datasets: MAPS, MAESTRO, and Studio MAESTRO. We evaluate Kong et al.'s and our best model trained with data augmentation. Kong's model suffers a significant drop in F1 score across all conditions, as much as 19.2 percentage points. Our model is mostly invariant to the perturbations, with the largest drop of only 2.5.

\subsection{Single Augmentations}
\label{sec:single_augmentations}
To better understand the influence of data augmentation during training, we conduct a series of training experiments during which we apply only one augmentation. Furthermore, we look at the influence of selection of training data, using either the original MAESTRO or the Studio MAESTRO re-recording. Each augmentation is applied during training time, again with a 50\% probability of being applied per example. Due to resource constraints, we limit the training to 28,000 steps (roughly 10 epochs over the ground truth labels). All other training parameters match those described in Section \ref{sec:results}. The results are shown in Table \ref{tab:single_aug}. The importance of pitch shifting and reverb as data augmentation techniques is evident, with out-of-distribution improvements in F1 score of 3.1\% and 2.8\% respectively. In these experiments, the additive background noise augmentation seems insignificant or even detrimental to performance. This is likely due to the clean recording conditions of the MAPS test set. The EQ augmentation also does not appear to help. This could indicate poor parameterization of the equalizer, or perhaps too weak boosts and cuts.

\begin{table}[htp]\centering
\caption{{\bf Single Augmentation Results:} \\ Each value corresponds to the note-onset F1 score when evaluating on the MAPS test set. Each experimental condition used a single data augmentation and either the original MAESTRO data or the Studio MAESTRO data.}\label{tab:single_aug}
\vspace{0.25cm}
\scriptsize
\begin{tabular}{lllll}\toprule
& &\multicolumn{2}{c}{Training data} \\\cmidrule{3-4}
&\textbf{} &MAESTRO &Studio MAESTRO \\\midrule
\multirow{5}{*}{Augmentation} &No Augmentation & 82.4 \tablefootnote{For this condition, we  report the evaluation of the published model.} & 75.2 \\ 
&Background &82.7 (+0.3) &73.3 (-1.9) \\
&Pitch Shift &85.5 (\textbf{+3.1}) &78.3 (\textbf{+3.1}) \\
&Reverb & 85.2 (\textbf{+2.8}) &76.6 (+1.4) \\
&EQ &82.1 (-0.3) & 74.4 (-0.8) \\
\bottomrule
\end{tabular}
\end{table}

\subsection{Ablation Study}
\label{sec:ablation_study}
In the next series of experiments, we again run training for 28,000 steps, but this time we remove only one component from the full data augmentation pipeline. The results are reported in Table \ref{tab:ablation}, with the first row being the full data augmentation pipeline. Once again, the strong influence of pitch shifting and reverb are demonstrated by the relatively large drop in performance when skipping these augmentations. Some interaction effects seem to be present between the various augmentations. In particular, the EQ augmentation appears inconsequential or even detrimental when applied as a single augmentation, but removing it from the pipeline in the ablation study hurts results.

%If the table is too wide, replace \begin{table}[!htp]...\end{table} with
%\begin{adjustwidth}{-2.5 cm}{-2.5 cm}\centering\begin{threeparttable}[!htb]...\end{threeparttable}\end{adjustwidth}
\begin{table}[!t]\centering
\caption{{\bf Ablation Experiments}: \\ In these experiments, we train the model with versions of the data augmentation pipeline with one augmentation removed. Note-onset F1 score on the MAPS test set is reported.}\label{tab:ablation}
\scriptsize
\begin{tabular}{lllll}\toprule
\textbf{} & &\multicolumn{2}{c}{Training data} \\\cmidrule{3-4}
& &MAESTRO &Studio MAESTRO \\\midrule
\multirow{5}{*}{Ablation} &Full Augmentation & 86.4 & 79.0 \\ 
&Skip Background &85.4 (-1.0) &78.8 (-0.2) \\
&Skip Pitch Shift &82.9 (\textbf{-3.5}) &75.6 (\textbf{-3.4}) \\
&Skip Reverb &82.8 (\textbf{-3.6}) &77.1 (-1.9) \\
&Skip EQ & 86.4 (-0.0) & 77.8 (-1.2) \\
\bottomrule
\end{tabular}
\end{table}

\section{Conclusion}
This work explored the role of data and data augmentation techniques in enhancing the robustness of APT systems, particularly regarding their performance on out-of-distribution data. Utilizing a new studio recording of the MAESTRO dataset and simple data augmentation, our research yields a new SOTA on the MAPS test set and new insights into the influence of data augmentation on piano transcription models.

Despite the strong performance on the MAPS test set, the findings underscore the nuanced and occasionally divergent effects of different data augmentations. Particularly, pitch shifting and reverb emerged as substantial in improving model generalization, while others like EQ augmentation did not yield anticipated benefits. A limitation of our approach is that the combinatorial nature of exploring mixtures of data augmentations made an exhaustive study prohibitively expensive. Furthermore, we were unable to conduct precise tests of significance, which would require running the training task for every condition multiple times and measuring variance across runs. Additional evaluation datasets are necessary to further measure the robustness of our model; in particular, live performances in a noisy environment would provide a better indication of whether the background noise augmentation is useful.

Moving forward, one avenue for future work lies in leveraging the advanced physical modeling features of Pianoteq to synthesize piano audio with diverse tunings and string inharmonicities, which could further enrich the timbral diversity of training data. Additionally, the studio recording of MAESTRO may present a valuable resource for other research areas, such as audio style transfer, offering a unique opportunity to delve into style translations between different recordings. The gap between performance on MAESTRO test and MAPS test motivates further research into the differences between these datasets, to ensure accuracy of their labels and identify any inconsistencies.

In light of our findings, we advocate for a heightened emphasis on out-of-distribution evaluation in future music transcription research, not only to benchmark model performance but also to delve deeper into understanding and enhancing model robustness and generalization across varied and potentially challenging acoustic environments.

\vfill\pagebreak

%\section*{Acknowledgements}
%
%The first author is a research student at the UKRI Centre for Doctoral Training in Artificial Intelligence and Music, supported jointly by UK Research and Innovation (grant number EP/S022694/1), Queen Mary University of London, and Yamaha Corporation.
%This work took place while the first author was performing a research internship at Yamaha Corporation.
%Additionally, this project made use of time on Tier 2 HPC facility JADE2, funded by the Engineering and Physical Sciences Research Council (EP/T022205/1).

%\vfill\pagebreak

\bibliographystyle{IEEEbib}
\bibliography{refs}

\end{document}